\documentclass{article}
\usepackage{jheppub}
\usepackage{amsmath}
\usepackage{amsfonts}
\usepackage{amssymb}
\usepackage{graphicx}

\def\be{\begin{equation}}

\def\ee{\end{equation}}
\def\bes{\begin{equation}\begin{split}&}
\def\es{\end{split}}
\def\bi{\bibitem}
\title{Early Universe with modified scalar-tensor theory of gravity.}

\author[a]{Ranajit Mandal}
\author[b]{Chandramouli Sarkar}
\author[c]{Abhik Kumar Sanyal}
\affiliation[a]{Dept. of Physics\\
                University of Kalyani, Nadia, India - 741235}
\affiliation[b]{Dept. of Physics \\
                Jangipur College, Murshidabad, India - 742213}
\affiliation[c]{Dept. of Physics \\
                Jangipur College, Murshidabad, India - 742213}

\emailAdd{ranajitmandalphys@gmail.com}
\emailAdd{csarkar123@yahoo.co.in}
\emailAdd{sanyal\_ ak@yahoo.com}

\abstract
{Scalar-tensor theory of gravity with non-minimal coupling is a fairly good candidate for dark energy, required to explain late-time cosmic evolution. Here we study the very early stage of evolution of the universe with a modified version of the theory, which includes scalar curvature squared term. One of the key aspects of the present study is that, the quantum dynamics of the action under consideration ends up generically with de-Sitter expansion under semiclassical approximation, rather than power-law. This justifies the analysis of inflationary regime with de-Sitter expansion. The other key aspect is that, while studying gravitational perturbation, the perturbed generalized scalar field equation obtained from the perturbed action, when matched with the perturbed form of the background scalar field equation, relates the coupling parameter and the potential exactly in the same manner as the solution of classical field equations does, assuming de-Sitter expansion. The study also reveals that the quantum theory is well behaved, inflationary parameters fall well within the observational limit and quantum perturbation analysis shows that the power-spectrum does not deviate considerably from the standard one obtained from minimally coupled theory.}


\begin{document}
\maketitle
\flushbottom

\section{Introduction}

In the early stage of evolution, between $10^{-36}$ sec - $10^{-32}$ sec, the universe went through an inflationary phase of expansion, which has been accepted as a scenario rather than a model. What happened prior to the inflationary era can only be answered in view of `quantum theory of gravity'. In the absence of a complete and legitimate (unitary and renormalizable) quantum theory of gravity, some perception regarding the evolution of the universe in the pre-inflationary  phase may be achieved through quantization of a viable cosmological model - known as ``quantum cosmology". In this regard a cosmological model should be chosen such that it leads to a viable inflationary era fitting the recently available data, and also the late-time cosmic acceleration which is a great cry of the present century. Non-minimally coupled scalar-tensor theory of gravity having an action in the form

\be\label{A1} A_1=\int\sqrt{-g}\;d^4x\left[f(\phi){R}-\frac{1}{2}\phi_{,\mu}\phi^{,\mu}-V(\phi)\right],\ee
is a strong contender for dark-energy candidature, being able to explain late-time cosmic acceleration inclusive of presently available data \cite{late}. It is therefore required to test the theory during early phase of cosmological evolution. To accommodate cosmic inflation in the theory, additional contribution to the above action is required. It is well known that quantum gravity, in any of its forms (super string theory, heterotic string theory super-gravity, M-theory, loop quantum gravity etc.) admits higher order curvature invariant terms. Of particular interest is curvature squared ($R^2, R_{\mu\nu}^2$) term, since it leads to renormalized theory of gravity \cite{stell}. Although, perturbatively it encounters ghost degrees of freedom, non-perturbatively it is well behaved. Further, although unification of the early inflation with late-time cosmic acceleration with non-minimally coupled scalar field has recently been achieved \cite{Beh}, nevertheless one can't ignore the presence of higher order terms in the early universe, particularly because inflation without phase transition is an essential feature of such higher order term \cite{staro, maeda}. Now, in the homogeneous and isotropic background, $R_{\mu\nu}^2$ and $R^2$ terms differ only by a total derivative term, and it is therefore sufficient to add scalar curvature squared term to the above action, which therefore reads as

\be\label{A2} A=\int\sqrt{-g}\;d^4x\left[f(\phi){R}+B R^2-\frac{1}{2}\phi_{,\mu}\phi^{,\mu}-V(\phi)\right].\ee

As already mentioned, our present aim is not to establish unification, rather to study its behaviour at the very early stage of cosmic evolution particularly, in the quantum domain. This requires canonical formulation of the theory. There exists a host of canonical formalisms in connection with higher order theory of gravity, which requires additional degrees of freedom. In addition to the three space metric ($h_{ij}$), the extrinsic curvature tensor ($K_{ij}$) is taken as an additional basic variable for the purpose. In the process, Cauchy data exceeds and therefore, one requires more boundary data. While, Ostrogrdski's technique \cite{ostro}, Dirac's \cite{dirac} formalism and Horowitz' no boundary proposal \cite{horo} insist upon setting $\delta h_{ij} = 0 = \delta K_{ij}$, at the boundary, and end up with the same phase-space Hamiltonian; a modified version of Horowitz' technique \cite{mod1, mod2, mod3, mod4, mod5} favours $\delta h_{ij} = 0 = \delta R$, at the boundary and for a class of higher order theory of gravity, ends up with a Hamiltonian which although different, is related under a suitable canonical transformation with the one obtained following the known standard techniques \cite{ostro, dirac, horo}. It has therefore been tacitly assumed that the canonical structure of higher order theory of gravity is independent of the choice of boundary condition. However, such a preconceived concept has shattered in view of a very recent work \cite{mod6}. It has been found that for Gauss-Bonnet-dilatonic coupled higher order theory of gravity,

\be A_G =\int\sqrt{-g}\;d^4x\left[\alpha{R}+\beta R^2+\Lambda(\phi)\mathcal{G}-\frac{1}{2}\phi_{,\mu}\phi^{,\nu}-V(\phi)\right] + \alpha\Sigma_R + \beta\Sigma_{R^2} + \Lambda(\phi)\Sigma_\mathcal{G},\ee
(where, $\Sigma_R = 2 \oint_{\partial\mathcal{V}}K \sqrt hd^3x$, $\Sigma_{R^2} = 4 \oint_{\partial\mathcal{V}} R K\sqrt h d^3x$ and $\Sigma_\mathcal{G} = 4\oint_{\partial\mathcal{V}} \left( 2G_{ij}K^{ij} + \frac{\mathcal{K}}{3}\right)\sqrt hd^3x$ are the supplementary boundary terms known as the Gibbons-Hawking-York term corresponding to the linear sector, its modified version for curvature squared term and for Gauss-Bonnet-dilatonic coupled sector respectively, $\alpha, \beta, ~\mathrm{and}~ \Lambda(\phi)$ are the coupling parameters, $V(\phi)$ is the dilatonic potential, while the symbol $\mathcal{K}$ stands for $\mathcal{K}=K^3 - 3K K^{ij}K_{ij} + 2K^{ij}K_{ik}K^k_j$, $K$ being the trace of extrinsic curvature tensor), modified Horowitz' formalism ends up with a different phase-space Hamiltonian, which is not related to the others under canonical transformation \cite{mod6}. The most important outcome of the work is that, the well-known standard formalisms \cite{ostro, dirac, horo}, for which supplementary boundary terms are not required due to the fact that $\delta h_{ij} = 0 = \delta K_{ij}$ at the boundary, don't produce correct classical analogue of the theory under appropriate semi-classical approximation, although modified Horowitz' formalism does \cite{mod6}. It is true that the choice of boundary terms does not in any way affect classical field equations, nevertheless, the above mentioned results prove unambiguously that it does, in the quantum domain, and the standard techniques \cite{ostro, dirac, horo} towards canonical formulation don't render a viable quantum description of the theory, in general. Here, we have chosen the action \eqref{A2} to establish in particular the fact that, the in-equivalent phase-space structure of the Hamiltonian is an outcome of non-minimal coupling.\\

It is important to repeat that, if one insists upon the boundary conditions $\delta h_{ij}|_{\mathcal{\partial V}}=0=\delta R|_{\mathcal{\partial V}}$,
the action must be supplemented by additional boundary terms. On the contrary, the boundary conditions $\delta h_{ij}|_{\mathcal{\partial V}}=0=
\delta K_{ij}|_{\mathcal{\partial V}}$ take care of the total derivative terms, and there is no need to supplement the action with additional boundary terms.
This was the main argument of Horowitz for his no boundary proposal. Horowitz \cite{horo} argued that, without the supplementary boundary term, superposition principle holds during the transition from the initial configuration space to the final, following an intermediate one. However, Horowitz \cite{horo} also pointed out that, the above argument does't specifically state that boundary terms can't exist. Although sounds attractive, with no boundary proposal the cherished Gibbons-Hawking-York \cite{boun1, boun2} term, which is responsible for the entire contribution to the Euclidean action, also vanishes, and it is not possible to recover it under weak field limit. Next, it is well-known that $F(R)$ theory of gravity admits scalar tensor equivalence, under redefinition of $F(R)$ by an auxiliary variable to Jordan's frame or through conformal transformation to Einstein's frame. Variation of such canonical Lagrangian requires to fix the scalar at the end point. This is indeed equivalent to fixing of the Ricci scalar $R$ at the boundary. Further, Dyer and Hinterbichler \cite{dyre} have shown that the boundary terms reproduce the expected ADM energy, and the entropy of a Schwarzschild black hole is one-quarter of the area of the horizon in units of the effective Planck's length, which agrees with the Wald entropy formula \cite{wald1, wald2}. This clearly indicates that higher curvature terms make no additional correction to the entropy in the result obtained from Gibbons-Hawking-York \cite{boun1, boun2} term. Last but not the least important fact, as already mentioned is that, the quantum counterpart with no boundary proposal for Gauss-Bonnet-dilatonic coupled higher order theory of gravity does not lead to a classical limit under an appropriate semi-classical approximation \cite{mod6}. \\

Usually, either exponential expansion in the de-Sitter form or power law expansion as a solution to the classical field equations, is the starting point of studying inflationary evolution. The aim of the present work is to explore the fact that for the action under consideration \eqref{A2}, the quantum domain generically leads to de-Sitter expansion under appropriate semi-classical approximation, rather than power law expansion. This is the most important outcome of the present work. We therefore make canonical formulation of the action \eqref{A2}, in the Robertson-Walker minisuperspace, canonically quantize and show that the semiclassical wave-function is strongly peaked around the classical de-Sitter solution. Next we show that such a de-Sitter solution goes through to a viable inflationary phase, and the inflationary parameters are very much consistent with currently available data.\\

In the following section we therefore follow modified Horowitz' technique towards canonical formulation of action \eqref{A2} in the homogeneous and isotropic background. However, we encounter certain operator ordering ambiguity during quantization, which can only be resolved after having specific knowledge on the form of the coupling parameter $f(\phi)$. We therefore choose de-Sitter solution, and find a relation between the coupling parameter $f(\phi)$ and the potential $V(\phi)$ in view of the $(^0_0)$ equation of Einstein. The other field equation then fixes the forms of the two, uniquely. In view of the form of the coupling parameter $f(\phi)$, we remove operator ordering ambiguity, present the quantum mechanical probabilistic interpretation, and perform semiclassical approximation to explore the already mentioned fact that, the wave-function is strongly peaked around the classical de-Sitter solution.\\

Since quantum dynamics of action \eqref{A2} admits de-Sitter solution as its generic feature, so in section 3, we study the inflationary regime, with exponential expansion. For this purpose, we first translate our action \eqref{A2} to the Einstein's frame under conformal transformation, to show that the present action involves an additional scalar field $\psi$. This clarifies the reason for imposing a condition in addition to the standard slow-roll approximation. However, the rest of the analysis has been performed in the original variables. While studying gravitational perturbation, we match perturbed generalized scalar-field equation with the perturbed background scalar-field equation and observe that the coupling parameter $f(\phi)$ and the potential $V(\phi)$ must be related exactly in the same manner, as has been found while solving classical field equations assuming de-Sitter exponential solution.
In appendix A, we prove that the effective canonical Hamiltonian is hermitian. In appendix B, we show that the standard canonical formulation schemes following Ostrogradski's \cite{ostro}, Dirac's \cite{dirac} and Horowitz' \cite{horo}, lead to the same phase-space structure of the Hamiltonian, which is different from the one obtained following modified Horowitz' technique in section 2, and is not related to it under canonical transformation.

\section{Canonical formulation of non-minimally coupled scalar-tensor theory of gravity in the presence of higher order term}

In accordance with the discussions in the introduction, we insist upon keeping $\delta h_{ij} = 0 = \delta R$, at the boundary, and so it is required to
supplement the action \eqref{A2} with appropriate boundary terms. The action \eqref{A2} therefore reads

\be\label{2.3} A=\int\sqrt{-g}\;d^4x\left[f(\phi){R}+B R^2-\frac{1}{2}\phi_{,\mu}\phi^{,\mu}-V(\phi)\right]+ \Sigma_R +\Sigma_{R^2_1}
+\Sigma_{R^2_2},\ee
where, $\Sigma_R = 2\int f(\phi)K\sqrt{h}~ d^3x$ is the modified Gibbons-Hawking-York boundary term in the presence of non-minimal coupling, and $\Sigma_{R^2} = \Sigma_{R^2_1} +\Sigma_{R^2_2}
= 4B\int^3R K\sqrt{h}$ is the boundary term associated with the scalar curvature square term ($R^2$), which has been split into two parts in the manner,
$\Sigma_{R^2_1}  = 4B\int~^3R K\sqrt{h} ~d^3x;\;\Sigma_{R^2_2} = 4B\int(^4R - ^3R) K\sqrt{h} ~d^3x$. Such splitting is required to take care of the total
derivative terms appearing under integration by parts. This has been discussed earlier in details \cite{mod1, mod2, mod3, mod4, mod5, mod6}. In the above, $K$ is the trace of the extrinsic curvature tensor, and $h$ is the determinant of the induced three-space metric. Since reduction of higher order theory to its canonical form requires additional degrees of freedom, hence, in addition to the three-space metric $h_{ij}$, the extrinsic curvature tensor $K_{ij}$ is treated as basic variable. In the homogeneous and isotropic Robertson-Walker metric, viz.,
\be\label{2.4} ds^2 = - N(t)^2 dt^2 + a^2(t) \left[\frac{dr^2}{1-kr^2} + r^2 (d\theta^2 + sin^2 \theta d\phi^2)\right],\ee
we therefore choose the basic variables $h_{ij} = z \delta_{ij}= a^2 \delta_{ij}$, so that $K_{ij} = -{\dot h_{ij}\over 2 N} = -{a\dot a\over N} \delta_{ij}= -\frac{\dot z}{2 N} \delta_{ij}$, and hence, the Ricci scalar is expressed as,

\be \label{2.5} R =\frac{6}{N^2}\left(\frac{\ddot a}{a}+\frac{\dot a^2}{a^2}+N^2\frac{k}{a^2}-\frac{\dot N\dot a}{N a}\right)={6\over N^2}\left({\ddot z\over 2z} + N^2 {k\over z} - {1\over 2}{\dot N\dot z\over N z}\right),\ee
The action \eqref{2.3} in Robertson-Walker minisuperspace \eqref{2.4}, therefore takes the form,

\be\begin{split}\label{2.6}
A &= \int\bigg[{3 f\sqrt z}\Big(\frac{\ddot z}{ N}- \frac{\dot N \dot z}{N^2} + 2k N \Big)+\frac{9B}{\sqrt z}\Big(\frac{{\ddot z}^2}{N^3} -
\frac{2 \dot N \dot z \ddot z}{N^4} + \frac{{\dot N}^2{\dot z}^2}{N^5} -\frac{4k\dot N \dot z}{N^2}+ \frac{4 k {\ddot z}}{N} + 4 k^2 N \Big)\\&\hspace{2.5 in}+z^{\frac{3}{2}} \Big(\frac{1}{2N}\dot\phi^2-VN\Big)\bigg]dt + \Sigma_R +\Sigma_{R^2_1} +\Sigma_{R^2_2}.
\end{split}\ee
In the above, $\Sigma_R =-\frac{3f\sqrt z\dot z}{N},~~\text{while}~~\Sigma_{R^2_1}= -\frac{36 B k\dot z}{N\sqrt z}~~\text{and}  ~~\Sigma_{R^2_2}
=-\frac{18B\dot z}{N^3\sqrt z}\left({\ddot z}-\frac{\dot z\dot N}{N} \right)$. Under integrating by parts, the counter terms $\Sigma_R$ and
$\Sigma_{R^2_1}$ get cancelled and the action (\ref{2.6}) reduces to
\be\begin{split}\label{2.7}
A &= \int\bigg[\Big(-\frac{3f'\dot\phi\dot{z} \sqrt z}{N} - \frac{3 f{\dot z}^2}{2 N\sqrt z} + 6 kN f \sqrt z \Big) + \frac{9 B}{\sqrt z}
\Big(\frac{{\ddot z}^2}{N^3}-\frac{2\dot N\dot z \ddot z}{N^4}+\frac{\dot N^2\dot z^2}{N^5} + \frac{2 k {\dot z}^2}{N} + 4 k^2N  \Big)\\& \hspace{3.5 in} +z^{\frac{3}{2}}\Big(\frac{1}{2N}\dot\phi^2-VN\Big)\bigg]dt + \Sigma_{R^2_2},
\end{split}\ee
where, prime denotes derivative with respect to the scalar field $\phi$. At this stage introducing an auxiliary variable
\be\label{2.8}\mathcal Q=\frac{\partial A}{\partial \ddot z}=\frac{18 B}{N^3\sqrt z}\left({\ddot z}
-\frac{\dot N\dot z}{N}\right)\ee
straight into the action (\ref{2.7}), as
\be\begin{split}\label{2.9}
A&= \int\Bigg[\left(-\frac{3f'\dot\phi\dot{z} \sqrt z}{N} - \frac{3 f{\dot z}^2}{2 N\sqrt z} + 6 k Nf \sqrt z \right)+ \mathcal Q\ddot z-
\frac{N^3\sqrt{z}}{36B}{\mathcal Q}^2-\frac{\dot N\dot z \mathcal{Q}}{N}+ \frac{18B k\dot z^2}{Nz^{\frac{3}{2}}}+\frac{36BNk^2}{\sqrt{z}} \\& \hspace{3.5 in} +z^{\frac{3}{2}}\left(\frac{1}{2N}\dot\phi^2-VN\right)\Bigg]dt+ \Sigma_{R^2_2},\end{split}\ee
the rest of the boundary terms is taken care of under integration by parts. Finally, the action being free from the boundary temrs, is expressed as
\be\begin{split}\label{2.10}
A &= \int\Bigg[-\dot {\mathcal Q}\dot z-\frac{3f'\dot\phi\dot{z} \sqrt z}{N} - \frac{3 f{\dot z}^2}{2 N\sqrt z}+ 6 k N f \sqrt z -
\frac{N^3\sqrt{z}}{36B}{\mathcal Q}^2-\frac{\dot N\dot z\mathcal{Q}}{N}+ \frac{18B k\dot z^2}{Nz^{\frac{3}{2}}}+\frac{36BNk^2}{\sqrt{z}} \\&
\hspace{3.8 in}+z^{\frac{3}{2}}\left(\frac{1}{2N}\dot\phi^2-VN\right)\Bigg]dt.\end{split}\ee
The canonical momenta are
\begin{subequations}\label{2.11}\begin{align}
& p_{\mathcal Q} = - \dot z \label{pq} \\
&p_z =-\dot {\mathcal Q} -\frac{3f'\dot\phi\sqrt z}{N} -\frac{3f\dot z}{N\sqrt{z}}-
\frac{\dot N \mathcal Q}{N}+\frac{36 B k \dot z}{N z^{\frac {3}{2}}} \label{pz} \\
& p_N=-\frac{\dot z \mathcal Q}{N}\\
& p_\phi = - \frac{3f' \dot z \sqrt z}{ N}+{{z^{\frac{3}{2}}}\dot\phi\over N}  \label{pphi}
\end{align}\end{subequations}
The $(^0_0)$ component of Einstein's field equation in terms of the scale factor is

\be\begin{split}\label{2.12}
~&\Bigg[-{6f\over a^2}\left({\dot a^2\over N^2}+k\right)-6{f'\dot a\dot\phi\over N^2 a} -{36B\over a^2N^4}\Bigg(2\dot a\dddot a -
\ddot a^2 +2{\dot a^2\ddot a\over a} - 3{\dot a^4\over a^2} - 2\dot a^2{\ddot N\over N} - 4{\dot N\over N} \dot a\ddot a + 5 \dot a^2{\dot N^2\over N^2}
- 2 {\dot a^3 \dot N\over a N}\\&\hspace{2.8 in} - 2k N^2{\dot a^2\over a^2} +k^2{N^4\over a^2}\Bigg) + \left({\dot \phi^2\over 2 N^2} + V(\phi)\right)\Bigg]Na^3=0, \end{split}\ee
which when expressed in terms of the phase-space variables leads to the Hamilton constraint equation. However, construction of the phase-space structure of the Hamiltonian is non-trivial, since the Hessian determinant vanishes and the Lagrangian corresponding to the action \eqref{2.10} is degenerate. This is due to the presence of the time derivative of the lapse (which is essentially a Lagrange multiplier of the theory) in the said action, which is a typical to the higher order theory. Remember that no such time-derivative of the lapse function appears in General Theory of Relativity. The constraint $Q p_Q - N p_N = 0$, is also apparent from the expressions of canonical momenta \eqref{2.11}. Usually Dirac's constraint analysis is invoked to construct the phase-space Hamiltonian. Nevertheless, we have repeatedly pointed out \cite{mod3} that it is possible to bypass the issue and construct the canonical Hamiltonian in the following manner. For this purpose, let us use the expression,
\be \label{2.13} p_{\mathcal Q} p_z = \dot {\mathcal Q}\dot z +\frac{3\dot z f'\dot\phi\sqrt z}{N} +\frac{3f\dot z^2}{N\sqrt{z}}+
\frac{\dot N\dot z \mathcal Q}{N}-\frac{36 B k \dot z^2}{N z^{\frac {3}{2}}}\ee
obtained in view of the relations (\ref{pq} and \ref{pz}), and construct the Hamiltonian constraint equation in terms of the phase space variables as,
\be\begin{split}\label{2.14}
H_c &= 3f\Big(\frac{{p_{\mathcal Q}}^2}{2 N \sqrt z} - 2k N \sqrt z \Big) - p_{\mathcal Q} p_z + \frac{N^3 {\mathcal Q}^2 \sqrt z}{36B} -
\frac{18 k B}{N\sqrt z} \Big(\frac{{p_\mathcal Q}^2}{ z} + 2kN^2 \Big) +\frac{Np_{\phi}^2}{2z^{\frac{3}{2}}}-\frac{3f'p_Qp_{\phi}}{z}\\&\hspace{3.9 in}+\frac{9f'^2p_Q^2}{2N\sqrt z}+VNz^{\frac{3}{2}}= 0.
\end{split}\ee
It is important to note that although the phase-space structure of the Hamiltonian corresponding to the higher order theory under consideration has been produced in \eqref{2.14}, it does not establish the diffeomorphic invariance of the theory. Diffeomorphic invariance is apparent only when the Hamiltonian is expressed in terms of the basic variables ($x, z, \phi$) and their canonically conjugate momenta ($p_x, p_z, p_{\phi}$). In order to express the Hamiltonian in terms of the basic variables, we make the canonical transformations following the replacements of
$\mathcal Q = \frac{p_x}{N}$ and $ p_{\mathcal Q} = -Nx$, as before. The phase-space structure of the Hamiltonian in terms of the basic variables is then expressed as,
\be\begin{split}\label{2.15}
H_c &= N\Big[x p_z + \frac{ \sqrt z {p_x}^2}{36B} + 3f\Big(\frac{x^2}{2 \sqrt z} - 2k \sqrt z \Big) - \frac{18 k B}{\sqrt z} \Big(\frac{x^2}{z} + 2k\Big)+
\frac{p_{\phi}^2}{2z^{\frac{3}{2}}}+\frac{3f'xp_{\phi}}{z}+\frac{9f'^2x^2}{2\sqrt z}+Vz^{\frac{3}{2}}\Big]\\&=N\mathcal{H}.
\end{split}\ee
Diffeomorphic invariance is now clearly established in the equation \eqref{2.15}. The action (\ref{2.7}) can also be expressed in the canonical form with respect to the basic variables as,
\be\begin{split}\label{2.16} A &= \int\left(\dot z p_z + \dot x p_x +\dot\phi p_{\phi}- N\mathcal{H}_L\right)dt~ d^3 x
= \int\left(\dot h_{ij} \pi^{ij} + \dot K_{ij}\Pi^{ij}+\dot\phi p_{\phi} - N\mathcal{H}_L\right)dt~ d^3 x,\end{split}\ee
where, $\pi^{ij}$ and $\Pi^{ij}$ are momenta canonically conjugate to $h_{ij}$ and $K_{ij}$ respectively. This establishes the importance of the use of appropriate basic variables, over other canonical ones. Now, the canonical Hamiltonian \eqref{2.15} may immediately be quantized to obtain,
\be \begin{split}\label{2.17}
\frac{i\hbar}{\sqrt z}\frac{\partial \Psi}{\partial z} = &-\frac{\hbar^2}{36B x}\left(\frac{\partial^2}{\partial x^2} + \frac{n}{x}\frac{\partial}{\partial x}
\right)\Psi -\frac{\hbar^2}{2xz^2}\frac{\partial^2 \Psi}{\partial \phi^2} + {3 \over z^{\frac{3}{2}}}\widehat{f' p_\phi}\\&\hspace{1.0 in}+ \left[\frac{3fx}{2 z} +\frac{9f'^2x}{2z}+\frac{Vz}{x} -{6kf\over x} - {18kBx\over z^2} - {36k^2B\over xz}\right]\Psi = \hat H_e\Psi,
\end{split}\ee
where, $n$ is the operator ordering index. In the above expression \eqref{2.17} due to the presence of coupling between $f'(\phi)$ and $p_{\phi}$, there still remains some operator ordering ambiguity, which may only be resolved after having specific knowledge regarding the form of $f(\phi)$.

\subsection{In search of a form of $f(\phi)$ and Canonical quantization}

The nonminimally coupled action (\ref{2.3}) can be re-expressed (apart from the supplementary boundary terms) as,
\be\label{3.18} A=\int\sqrt{-g}\;d^4x\left[\frac{1}{2}\mathrm{f}(\phi,R)-\frac{1}{2}\phi_{,\mu}\phi^{,\mu}-V(\phi)\right],\ee
where $\mathrm{f}(\phi,R)=2f(\phi){R}+2B R^2$.  The field equations in the homogeneous and isotropic flat ($k=0$) Robertson-Walker metric background (with $N=1$), are expressed as

\be\label{3.22} H^2=\frac{1}{3F}\left(\frac{1}{2}\dot\phi^2+\frac{R F-\mathrm{f}+2V}{2}-3H\dot F\right),\ee
\be\label{3.23} \dot H=-\frac{1}{2F}\left(\dot\phi^2+\ddot F-H\dot F\right),\ee
\be\label{3.24} \ddot\phi+3H\dot\phi+\frac{1}{2}\left(2V'-\mathrm{f}'\right)=0,\ee
where $F = {\partial \mathrm{f}\over\partial R}$, $H=\frac{\dot a}{a}$, $R=6(\dot H+2H^2)$, and the scalar field equation (\ref{3.24}) follows from equations (\ref{3.22}) and (\ref{3.23}).
It is already known that inflation is an essential feature of higher order curvature term \cite{staro, maeda}. Here our aim, as already mentioned in the introduction, is to check if the quantum dynamics of the higher-order theory under consideration in the Trans-Planckian era, leads generically to the inflationary scenario in the post Planck's era. Therefore, we seek inflationary solution of the classical field equations \eqref{3.22} - \eqref{3.24} in the following standard de-Sitter exponential form

\be\label{3.65} a=a_0 e^{{\mathrm H}t},~~~\phi=\phi_0 e^{-{\mathrm H}t},\ee
where, constant `$\mathrm{H}$' is the Hubble parameter `$H$' in the de-Sitter regime, and $R = 12 \mathrm{H}^2$. Equation \eqref{3.24} relates $V'(\phi)$ and $\mathrm{f'(\phi,R)}$ and in the process $V(\phi)$ and $f(\phi)$ in the following manner,

\be\label{Rel1} 2V' - \mathrm{f'} = 4\mathrm{H}^2\phi \Longrightarrow V(\phi) - 12\mathrm{H}^2 f(\phi) = \mathrm{H}^2\phi^2 + c,\ee
where $c$ is an integration constant. In view of equation \eqref{3.22}, one can also find

\be\label{Rel2} 2V = 12 \mathrm{H}^2 f(\phi) - 12 \mathrm{H}^2 f'(\phi)\phi -\mathrm{H}^2\phi^2.\ee
The two equations \eqref{Rel1} and \eqref{Rel2} are simultaneously satisfied restricting the potential and the coupling parameter to the following forms

\be\label{3.66} V(\phi)=V_1+\frac{V_0}{\phi},\;\;\;\text{and}\;\;\;f(\phi)=f_0+\frac{f_1}{\phi}-\frac{\phi^2}{12},\ee
where $c = -f_0R - 2B R^2$, and the parametric constants, $V_0,\;V_1,\; f_0,\;\mathrm{and}\;f_1$ are related in the following manner,

\be \label{3.67}V_0=12f_1 \mathrm{H}^2,\;\;\; V_1 = 6f_0 \mathrm{H}^2.\ee
Since the form of the coupling parameter $f(\phi)$ has been explored in \eqref{3.66}, we can now proceed with our left out task regarding  canonical quantization. Choosing suitable operator ordering between $f'(\phi)$ and $p_{\phi}$, equation \eqref{2.17} now takes the form

\be\begin{split}\label{4.71}
\frac{i\hbar}{\sqrt z}\frac{\partial \Psi}{\partial z} &= -\frac{\hbar^2}{36B x}\left(\frac{\partial^2}{\partial x^2} + \frac{n}{x}\frac{\partial}{\partial x}
\right)\Psi - \frac{\hbar^2}{2x z^2}\frac{\partial^2\Psi}{\partial \phi^2} +\frac{i\hbar}{2z^{\frac{3}{2}}}\left(\frac{\phi^3+6f_1}{\phi^2}\right)\frac{\partial\Psi}{\partial\phi}+\frac{i\hbar}{4z^{\frac{3}{2}}}
\left(\frac{\phi^3-12f_1}{\phi^3} \right)\Psi\\&+  \left[\frac{3 x}{2z}\left(f_0+\frac{f_1}{\phi}-\frac{\phi^2}{12}\right)+\frac{9x}
{2z}\left(\frac{f_1}{\phi^2}+\frac{\phi}{6}\right)^2
+\frac{6f_0 \mathrm{H}^2z}{x}+\frac{12f_1\mathrm{H}^2z}{x\phi}\right]\Psi ,\end{split}\ee
where Weyl symmetric ordering has been performed between $f'(\phi)$ and $p_\phi$, appearing in the third term on the right hand side, and the form of the potential has been supplemented from equation \eqref{3.66} and \eqref{3.67}. Now, under a change of variable, the above modified Wheeler-de-Witt equation, takes the look of Schr\"odinger equation, viz.,

\be\begin{split}\label{4.72}
i\hbar\frac{\partial \Psi}{\partial \sigma} &= -\frac{\hbar^2}{54B}\left(\frac{1}{x}\frac{\partial^2}{\partial x^2} + \frac{n}{x^2}\frac{\partial}{\partial x}
\right)\Psi - \frac{\hbar^2}{3x \sigma^{\frac{4}{3}}}\frac{\partial^2\Psi}{\partial\phi^2}+\frac{i\hbar}{3\sigma}\left(\frac{\phi^3+6f_1}{\phi^2}
\right)\frac{\partial\Psi}{\partial\phi}\\& \hspace{1.8 in}+\frac{i\hbar}{6\sigma}\left(\frac{\phi^3-12f_1}{\phi^3} \right)\Psi+  V_e\Psi  = \hat H_e\Psi,
\end{split}\ee
where, $\sigma  = z^{\frac{3}{2}} = a^3$ plays the role of internal time parameter. In the above equation, the effective potential $V_e$, is given by,

\be\label{4.73} V_e = \left[\frac{ x}{\sigma^{\frac{2}{3}}}\left(f_0+\frac{f_1}{\phi}-\frac{\phi^2}{12}\right)+\frac{3x}{\sigma^
{\frac{2}{3}}}\left(\frac{f_1}{\phi^2}+\frac{\phi}{6}\right)^2 +\frac{4f_0 \mathrm{H}^2\sigma^{\frac{2}{3}}}{x}+\frac{8f_1 \mathrm{H}^2\sigma^{\frac{2}{3}}}{x\phi}\right]. \ee

\subsection{Probabilistic interpretation}

The hermiticity of $\hat H_e$ (see Appendix A) should enable us to write the continuity equation, which requires to find $\frac{\partial\rho}{\partial\sigma}$, where,
$\rho=\Psi^*\Psi.$ A little bit of algebra leads to the following equation,
\be\label{4.74}\begin{split} \frac{\partial\rho}{\partial\sigma} &=-\frac{\partial}{\partial x}\left[\frac{i \hbar }{54B x}(\Psi\Psi^*_{,x}-\Psi^*\Psi_{~,x})
 \right]- \frac{\partial}{\partial \phi}\left[\frac{i \hbar}{3x\sigma^{\frac{4}{3}}}(\Psi\Psi^*_{,\phi}-\Psi^*\Psi_{~,\phi}) -\left(\frac{\phi^3+6f_1}{3\sigma\phi^2}\right)\Psi^*\Psi \right]\\&+\frac{(n+1)}{x^2}
 \left(\Psi\Psi^*_{,x}-\Psi^*\Psi_{,x}\right) \end{split}\ee
Clearly, continuity equation can be written, only under the choice $n = -1$, as
\be\label{4.75}\frac{\partial\rho}{\partial \sigma} + \nabla . {\bf{J}} = 0, \ee
where, $ \rho = \Psi^*\Psi ~~ \text{and} ~~  {\bf J} = ({\bf J}_x, {\bf J}_\phi, 0) $ are the probability density and the current density respectively, and

\begin{subequations}\label{4.76}\begin{align}
{\bf J}_x &= \frac{i \hbar }{54B x}(\Psi\Psi^*_{,x}-\Psi^*\Psi_{~,x})\\
{\bf J}_\phi &= \frac{i \hbar }{3x\sigma^{\frac{2}{3}}}(\Psi\Psi^*_{,\phi}-\Psi^*\Psi_{~,\phi})-\left(\frac{\phi^3+6f_1}{3\sigma\phi^2}\right)\Psi^*\Psi.\end{align}
\end{subequations}
It is noticeable that the existence condition of standard quantum mechanical probabilistic interpretation, fixes the operator ordering index to $n = -1$.

\subsection{Semiclassical approximation}

Now to check the viability of the quantum equation (\ref{4.72}), it is required to test its behaviour under certain appropriate semi-classical approximation.
For the purpose, it is easier to handle equation (\ref{4.71}) instead, and express it as,
\be\begin{split}\label{4.77}
-\frac{\hbar^2\sqrt z}{36B x}\left(\frac{\partial^2}{\partial x^2} + \frac{n}{x}\frac{\partial}{\partial x}\right)\Psi - \frac{\hbar^2}{2x z^
{\frac{3}{2}}}\frac{\partial^2\Psi}{\partial \phi^2} - i\hbar\frac{\partial \Psi}{\partial z}+\frac{i\hbar}{2z}\left(\frac{\phi^3+6f_1}{\phi^2}\right) \frac{\partial\Psi} {\partial\phi} + \mathcal{V}\Psi = 0
\end{split}\ee
where
\be\label{4.78} \mathcal{V} = \left[\frac{3 x}{2\sqrt z}\left(f_0+\frac{f_1}{\phi}-\frac{\phi^2}{12}\right)+\frac{9x}{2\sqrt z}\left(\frac{f_1}{\phi^2}+
\frac{\phi}{6}\right)^2 +\frac{6f_0{\mathrm{H}}^2z^{\frac{3}{2}}}{x}+\frac{12f_1{\mathrm{H}}^2z^{\frac{3}{2}}}{x\phi}+ \frac{i\hbar}{4z}\left(\frac{\phi^3-12f_1}{\phi^3} \right)\right] . \ee
The above equation \eqref{4.77} may be treated as time independent Schr{\"o}dinger equation with three variables ($x$, $z$, $\phi$). Therefore, as usual, let us seek the solution of equation (\ref{4.77}) as,

\be\label{4.79}\Psi = \psi_0e^{\frac{i}{\hbar}S(x,z,\phi)}\ee
and expand $S$ in power series of $\hbar$ as,
\be\label{4.80} S = S_0(x,z,\phi) + \hbar S_1(x,z,\phi) + \hbar^2S_2(x,z,\phi) + .... .\ee
Now inserting the expression (\ref{4.80}) in equation (\ref{4.79}) and taking appropriate derivatives of $\Psi$, everything may finally be inserted in equation (\ref{4.77}). At the end, equating the coefficients of different powers of $\hbar$ to zero, one obtains the following set of equations (up-to second order)
\begin{subequations}\begin{align}
&\frac{\sqrt z}{36B x}S_{0,x}^2 + \frac{S_{0,\phi}^2}{2xz^{\frac{3}{2}}} + S_{0,z}-\left(\frac{\phi^3+6f_1}{2z\phi^2}\right)S_{0,\phi} + \mathcal{V}(x,z,\phi) = 0, \label{4.81a} \\
&-\frac{i\sqrt z}{36B x}S_{0,xx} - \frac{in\sqrt z}{36B x^2}S_{0,x} - \frac{iS_{0,\phi\phi}}{2xz^{\frac{3}{2}}} + S_{1,z} + \frac{\sqrt zS_{0,x}S_{1,x}}{18Bx}
 + \frac{S_{0,\phi}S_{1,\phi}}{xz^{\frac{3}{2}}}-\left(\frac{\phi^3+6f_1}{2z\phi^2}\right)S_{1,\phi} = 0, \label{4.81b}\\
&-i\frac{\sqrt z S_{1,xx}}{36B x} - i\frac{n\sqrt zS_{1,x}}{36B x^2}+{\sqrt z}\frac{S_{1,x}^2+2 S_{0,x}S_{2,x}}{36B x} + \frac{ S_{1,\phi}^2+2S_{0,\phi}S_{2,\phi}}{2xz^{\frac{3}{2}}} -
i\frac{S_{1,\phi\phi}}{xz^{\frac{3}{2}}} + S_{2,z}-\left(\frac{\phi^3+6f_1}{2z\phi^2}\right)S_{2,\phi}= 0,\label{4.81c}
\end{align}\end{subequations}
that are to be solved successively to find $S_0(x,z,\phi),\; S_1(x,z,\phi)$ and $S_2(x,z,\phi)$ and so on. Now identifying $S_{0,x}$ as $p_x$; $S_{0,z}$ as
$p_z$ and $S_{0,\phi}$ as $p_{\phi}$ one can recover the classical Hamiltonian constraint equation $H_c = 0$, given in equation (\ref{2.15}) from equation
\eqref{4.81a}. This is a consistency check. Thus, $S_{0}(x, z)$ can now be expressed as,

\be\label{4.82} S_0 = \int p_z dz + \int p_x dx + \int p_\phi d\phi \ee
apart from a constant of integration which may be absorbed in $\psi_0$. The integrals in the above expression can be evaluated using the classical
solution for $k = 0$ presented in equation (\ref{3.65}), the definition of $p_z$ given in (\ref{pz}), $p_{\phi}$ in (\ref{pphi}) and $p_x = {\mathcal Q}$.
Further, we recall the expression for ${\mathcal Q}$ given in (\ref{2.8}), remember the relation, $x = \dot z$, where, $z = a^2$, choose
$n = -1$, since probability interpretation holds only for such value of $n$, and use the form of $f({\phi})$ presented in (\ref{3.66}), to obtain the following expressions of $p_x$, $p_z$ $p_\phi$ in term of $x$, $z$ and $\phi$,
\begin{subequations}\begin{align}
&\label{4.83} f'=-\left(\frac{f_1}{\phi^2}+\frac{\phi}{6}\right)\\
&x = 2{\mathrm H} z \\
&p_x = 36\sqrt 2 B {\mathrm H}^{\frac{3}{2}} \sqrt {x}\\
&p_z = -72B{\mathrm H}^3\sqrt z-6f_0\mathrm H \sqrt z-\frac{9f_1{\mathrm H}}{a_0\phi_0}z\\
&p_\phi = \frac{6f_1{\mathrm H} a_0^3\phi_0^3}{\phi^5}
                                \end{align}\end{subequations}
Hence the integrals in (\ref{4.82}) are evaluated as,
\begin{subequations}\begin{align}
&\label{4.84}\int p_x dx = 24 \sqrt{2} B {\mathrm H}^{\frac{3}{2}} x^{\frac{3}{2}}; \\
&\int p_z dz =  -48B{\mathrm H}^3z^{\frac{3}{2}}-4f_0\mathrm H z^{\frac{3}{2}}-\frac{9f_1 \mathrm H}{2a_0\phi_0}z^2; \\
&\int p_\phi d\phi = -\frac{3f_1{\mathrm H} a_0^3\phi_0^3}{2\phi^4}.
\end{align}\end{subequations}
\noindent
Thus, the explicit form of $S_0$ in terms of $z$ is found as,
\be\label{4.85}S_0 = \left(48B\mathrm{H}^3-4f_0\mathrm H\right) z^{\frac{3}{2}} -\frac{6f_1\mathrm H }{a_0\phi_0}z^2.\ee
For consistency, one can trivially check that the expression for $S_0$ (\ref{4.85}) so obtained, satisfies equation (\ref{4.81a}) identically. In fact
it should, because, equation (\ref{4.81a}) coincides with Hamiltonian constraint equation (\ref{2.15}) for $k = 0$. Moreover, one can also compute the zeroth
order on-shell action (\ref{2.10}). Using classical solution (\ref{3.65}) one may express all the variables in terms of $t$ and substitute in the action (\ref{2.10}) to obtain
\be\label{4.86}A=A_{cl}=\int\left[144B{\mathrm H}^4a_0^3e^{3{\mathrm H}t}-12f_0{\mathrm H}^2a_0^3e^{3{\mathrm H}t}-\frac{24f_1{\mathrm H}^2}{\phi_0}a_0^3
e^{4{\mathrm H}t}\right]dt. \ee
Integrating we have, \be\label{4.87} A=A_{cl}=\left(48B{\mathrm H}^3-4f_0{\mathrm H}\right)a_0^3e^{3{\mathrm H}t}-\frac{6f_1{\mathrm H}}{a_0\phi_0}a_0^4e^{4{\mathrm H}t},\ee
which is the same as we obtained in (\ref{4.85}), proving $S_0$ to be the classical action and checking consistency yet again. Hence, at this end, the wave function is
\be\label{4.88} \Psi = \psi_0 e^{\frac{i}{\hbar}\left[ \left(48B\mathrm{H}^3-4f_0\mathrm H\right) z^{\frac{3}{2}} -\frac{6f_1\mathrm H }
{a_0\phi_0}z^2\right]}.\ee

\subsubsection{ First order approximation}
Now for $n=-1$, equation \eqref{4.81b} can be expressed as,
\be\label{4.89} -\frac{\sqrt z}{36B x}\left(i S_{0,xx} - 2S_{0,x}S_{1,x} - \frac{i}{x}S_{0,x}\right) -\frac{1}{2 x z^{\frac{3}{2}}}\left(i S_{0,\phi\phi} - 2S_{0,\phi} S_{1,\phi}+\frac{x\sqrt z}{\phi^2}\left(\phi^3+6f_1\right)S_{1,\phi}\right) + S_{1,z}  = 0. \ee
\noindent
Using the expression for $S_0$ obtained in (\ref{4.85}), we can write $S_{1,z}$ from the above equation as
\be\label{4.90} S_{1,z} = {i \left[{12D\over a_0^2\phi_0^2}- {30\sqrt z\over a_0^3\phi_0^3} - {D\over 96B\mathrm{H}^2 z}\right] \over \left[2 +{D\over 24B \mathrm{H}^2}- {\sqrt z\over 12BH^2a_0\phi_0} +{12D z\over a_0^2\phi_0^2} + (6f_1 - 24){z^{3\over 2}\over a_0^3\phi_0^3}\right]},
\ee
where, $D = 12 B\mathrm{H}^2 -1$. On integration the form of $S_1$ is
found as,
\be \label{4.91}S_1 = i F(z),\ee
Therefore, the wavefunction up to first-order approximation reads
\be\label{4.92} \Psi = \psi_{01} e^{\frac{i}{\hbar}\left[ \left(48B\mathrm{H}^3-4f_0\mathrm H\right) z^{\frac{3}{2}} -
\frac{6f_1\mathrm H }{a_0\phi_0}z^2\right]},\ee
where,
\be\label{4.93}\psi_{01} = \psi_0 e^{-F(z)}.\ee
It has been proved that $S_0$ obeys Hamilton-Jacobi equation. Comparison with classical constraint equation $H_c = 0$ \eqref{2.15}, one finds $p_x = {\partial S_0\over \partial x}$, and $p_z = {\partial S_0\over \partial z}$. So the wavefunction shows a strong correlation between coordinates and momenta. Now using the relation between velocities and momenta and the fact that $S_0$ obeys Hamilton-Jacobi equation, it is apparent that the above relations define a set of trajectories in the $x-z$ plane, which are solutions to the classical field equations. Thus the semiclassical wave function \eqref{4.92} is strongly peaked around classical inflationary solutions \eqref{3.65}.

\section{Inflation}

Since the very early quantum universe smoothly transits to the de-Sitter type inflationary era, we therefore in the following subsection (3.1), compute the inflationary parameters in connection with the action \eqref{2.3} under consideration. It is important to note that the potential \eqref{3.66} is flat when $\phi$ is sufficiently large, and so, the slow-roll approximation is admissible. In the next subsection (3.2), we study classical aspect of gravitational perturbation. Finally in subsection (3.3) we derive the perturbation spectra generated from the quantum fluctuations in an early scalar field dominated accelerating (inflation) phase.

\subsection{Slow roll approximation}

Before we proceed, let us perform conformal transformation to demonstrate the fact that the higher order theory of gravity with non-minimal coupling under consideration \eqref{3.18}, involves an additional degree of freedom. Under the following conformal transformation,

\be \label{3.19} g_{\mu\nu}\rightarrow \hat{g}_{\mu\nu}=\Omega^2 g_{\mu\nu} ~\mathrm{and~under~the~choice}~ \Omega^2 =  F \equiv \frac{\partial{\mathrm{f}(\phi,R)}}{\partial R} = e^{\sqrt{\frac{2}{3}}\psi}\ee
where $\psi = \sqrt{3\over 2}\ln F$ is the new dynamical variable, the Lagrangian density, associated with action (\ref{3.18}) may be transformed into \cite{Hwang1, Hwang2}

\be \label{3.20} \hat L = \frac{1}{2}\hat R-\frac{1}{2 F}{\phi}_{\hat;\mu}{\phi}^{\hat;\nu}-\frac{1}{2}{\psi}_{\hat;\mu}{\psi}^{\hat;\nu}-\hat {V}(\phi,\psi), \;\;\;\mathrm{where,}\;\;\;
\hat {V}(\phi,\psi)=\frac{R F-\mathrm{f}+2V}{2F^2}.\ee
In the above, we use hats to denote quantities based on the conformally transformed metric frame. Thus, our original $\mathrm{f}(\phi,R)$ gravity is cast into the Einstein theory with an additional scalar field $\psi$, and a special potential term $\hat{V}(\phi,\psi)$ \eqref{3.20} \footnote{Only in a very special case, assuming $\psi = \psi(\phi)$, one can introduce another scalar field $\hat\phi$ to express the Lagrangian in terms of a single scalar field as in the case of minimally coupled scalar-tensor theory of gravity as, $\hat L = {\hat R\over 2} - {1\over 2} \hat{\phi}_{\hat;\mu}\hat{\phi}^{\hat;\mu} - \hat V(\hat \phi)$. However in that case, $\hat\phi$ should satisfy the relation $ d\hat\phi=\sqrt{\frac{1}{F}{d\phi}^2+{d\psi}^2}.$}. Using the conformal equivalence between the theories, it is possible to derive the equations for the background and the perturbations. The general asymptotic solutions for the perturbations in the generalized gravity from the simple results known in the minimally coupled scalar field may also be presented \cite{Hwang1, Hwang2}.
However, we shall not use the conformal transformation properties in the following treatment. Rather we use the conformal transformation to realize that, along with the standard slow-roll conditions of minimally coupled single-field inflation, viz. $\dot\phi^2\ll V$ and $|\ddot\phi|\ll3H|\dot\phi|$, it is required to impose one additional condition due to the presence of additional field viz. $4|\dot{\mathrm{f}}|H\ll 1$ \cite{addic}. As in the case of Gauss-Bonnet coupling \cite{hgb, hgb1, hgb2, hgb3}, instead of standard slow roll parameters, here also the introduction of a combined hierarchy in the following manner, appears to be much suitable. Firstly, the background evolution is described by a set of horizon flow functions (the behaviour of Hubble distance during inflation) starting from
\begin{center}
    \be\label{3.27}\epsilon_0=\frac{d_H}{d_{H_i}}, \ee
\end{center}
where, $d_H=H^{-1}$ is the Hubble distance, also called horizon in our chosen unit. We use suffix $i$ to denote the era at which inflation was initiated. Now hierarchy of functions is defined in a systematic way as
\be\label{3.28} \epsilon_{l+1}=\frac{d\ln|\epsilon_l|}{dN},~~~~~ l\geq 0.\ee
In view of the definition $N=\ln{\frac{a}{a_i}},$ which implies $\dot N=H,$ one can compute $\epsilon_1=\frac{d\ln{d_H}}{dN},$ which is the logarithmic
change of Hubble distance per e-fold expansion $N$. It is the first slow-roll parameter $\epsilon_1=\dot{d_H}=-\frac{\dot H}{H^2}$, implying that the Hubble parameter almost remains constant during inflation. The above hierarchy allows one to compute $\epsilon_2=\frac{d\ln{\epsilon_1}}{dN}=\frac{1}{H}\frac{\dot\epsilon_1}{\epsilon_1},$ which implies $\epsilon_1\epsilon_2=d_H\ddot{d_H}
=-\frac{1}{H^2}\left(\frac{\ddot H}{H}-2\frac{\dot H^2}{H^2}\right).$ In the same manner higher slow-roll parameters may be computed. Equation (\ref{3.27})
essentially defines a flow in space with cosmic time being the evolution parameter, which is described by the equation of motion

\be\label{3.29}\epsilon_0\dot\epsilon_l-\frac{1}{d_{H_i}}\epsilon_l\epsilon_{l+1}=0,~~~~l\geq 0.\ee
One can also check that (\ref{3.29}) yields all the results obtained from the hierarchy defined in (\ref{3.28}), using the definition (\ref{3.27}).
As already mentioned, the additional degree of freedom appearing due to the function $\mathrm{f}(\phi, R)$, requires to introduce yet another hierarchy of flow parameters as
\be\label{3.30} \delta_1=4\dot{\mathrm{f}} H\ll 1, ~~~~~\delta_{m+1}=\frac{d\ln|\delta_m|}{d\ln a}, ~~~~\text{with,} ~~~~m\geq 1.\ee
Clearly, for $m=1, \delta_2=\frac{d\ln|\delta_1|}{dN}=\frac{1}{\delta_1}\frac{\dot\delta_1}{\dot N},$ and $\delta_1\delta_2=\frac{4}{H}\left(\ddot{\mathrm{f}} H+\dot{\mathrm{f}}\dot H\right),$ and so on. The slow-roll conditions therefore read $|\epsilon_m|\ll 1$ and $|\delta_m| \ll 1$, which are analogous to the standard
slow-roll approximation. Now we arrange the field equations (\ref{3.22}) and (\ref{3.24}) in the following manner

\be\begin{split}\label{3.25} \dot\phi^2+2V&= 12 H^2 \mathrm{f}  + 3(1 + 4 H \dot{\mathrm{f}}) +144 B H^4\left(2+\frac{\dot H}{H^2}\right)- 72B H^4\left(2+\frac{\dot H}{H^2}\right)^2 \\&\hspace{0.7 in}- 144B H^4\left[1-\frac{1}{H^2}
\left(\frac{\ddot H}{H}-2\frac{\dot H^2}{H^2}\right)-2\left(1+\frac{\dot H}{H^2}\right)^2 + 1 \right] -3.
\end{split}\ee
\be\label{3.26} \ddot\phi+3H\dot\phi=-V'+ 6 H^2 \mathrm{f}'\left(2 + {\dot H\over H^2}\right) .\ee
In view of the slow-roll parameters, the above equations (\ref{3.25}) and (\ref{3.26}) may therefore be expressed as

\be\begin{split} \label{3.31} \dot\phi^2+2V& = 12 H^2\mathrm{f} + 3\left(1+\delta_1\right) +144 BH^4\left(2-\epsilon_1\right)-72 B H^4\left(2-\epsilon_1\right)^2
\\&-144 BH^4 \mathrm{f}\big[(1+\epsilon_1\epsilon_2)-2\left(1-\epsilon_1\right)^2 + 1\big]-3,\end{split}\ee
\be\label{3.32} \ddot\phi+3H\dot\phi=-V'+6 H^2 \mathrm{f}'(2-\epsilon_1),\ee
respectively, and may be approximated to

\be\label{3.33}V \simeq 6H^2 \mathrm{f},\ee
\be\label{3.34} 3H\dot\phi\simeq -V'+ 12 H^2 \mathrm{f}'. \ee
Since the Hubble parameter $H = \mathrm{H}$, remains constant during inflation, so the above pair of equations \eqref{3.33} and \eqref{3.34} may be combined to obtain

\be \label{3.35} \dot\phi={V'\over 3H}.\ee
The number of e-folds then has to be computed in view of the following relation,

\be\label{3.36} N(\phi)\simeq \int_{t_i}^{t_f}Hdt=\int_{\phi_i}^{\phi_f}\frac{H}{\dot\phi}d\phi\simeq \int_{\phi_i}^{\phi_f}
\frac{3H^2}{V'}d\phi,\ee
where, $\phi_i$ and $\phi_f$ denote the values of the scalar field at the beginning $(t_i)$ and the end $(t_f)$ of inflation. Since we already have knowledge on the specific forms of the potential $V(\phi)$ and the coupling parameter $f(\phi)$ in view of relations \eqref{3.66} and \eqref{3.67} it is now possible to compute the number of e-folding, along with the other slow-roll parameters.\\

\noindent
The number of e-folding (\ref{3.36}) now reads

\be\label{3.68}N(\phi)={1\over 4 f_1}\int_{\phi_f}^{\phi_i}\phi^2 d\phi=  {1\over 12 f_1}(\phi_i^3-\phi_f^3).\ee
We take following numerical values: $\phi_i = 4.8 M_p,\;\;\phi_f = 1.2 M_p,\;\; f_0 = 0.4291 M_p^2,\;\;\mathrm{and},\;\, f_{1} = -0.162 M_p^3$ to find that inflation halts ($\epsilon_f =1$) after $N  = 56$~e-fold of expansion.
The slow-roll parameters take the numerical values,
\be\label{3.69}\epsilon =\frac{M_{pl}^2}{2}\left(\frac{V'}{V}\right)^2=\frac{M_{pl}^2}{2}\left(\frac{V_0^2}{\phi^2
 \left(V_0+V_1\phi\right)^2}\right) = {M_p^2 \over \phi_i^2} \left({2f_1^2\over(2f_1 + f_0 \phi_i)^2}\right) = 0.0007562,\ee

\be\label{3.70}\eta = M_{pl}^2\frac{V''}{V}= M_{pl}^2 \left(\frac{2V_0}{\phi^2(V_0+V_1\phi)}\right) = {M_p^2 \over \phi_i^2}\left({4f_1 \over (2f_1 + f_0 \phi_i)}\right) = - 0.0162,\ee
and therefore the scalar to tensor ratio and the spectral index take the values $r = 16\epsilon = 0.012;$\;and,  $n_s = 1- 6\epsilon + 2\eta = 0.963$, which are very much within the limit of recently released data \cite{pd}.

\subsection{Gravitational perturbation: Validating the relation between $\mathrm{f}(\phi)$ and $V(\phi)$}

In this subsection, we study gravitational perturbation essentially to demonstrate the fact that the perturbed form of the background scalar field equation when equated to the same equation obtained by varying the metric-perturbed action , the relationship between the potential $V(\phi)$ and $\mathrm{f}(\phi, R)$ and hence between $V(\phi)$ and the coupling parameter $f(\phi)$ is found to be the same as already obtained \eqref{Rel1} in view of de-Sitter solution \eqref{3.65} of the classical field equations \eqref{3.22} and \eqref{3.24}. Although, it might appear striking, nevertheless, it validates the semiclassical approximation performed in subsection (2.3), that ends up with a wave-function which is oscillatory about de-Sitter expansion.\\

Cosmological perturbation in Einstein's gravitational theory \cite{Hwang3, Hwang4} and also in a broad class of theories $\mathrm{f}(\phi, R)$, often called generalized gravity theories, have been analysed thoroughly in a series of articles \cite{Hwang1, Hwang2, Hwang5, Hwang6, Hwang7, Hwang8, Noh}, in uniform curvature gauge, first introduced by Mukhanov \cite{Muk}. The asymptotic solutions of the scalar field equation in the large and small scale limits have been found. Further, the primordial seed density fluctuations generated from vacuum quantum fluctuations in different inflationary models for generalized gravity theories, have also been presented \cite{Hwang3, Hwang2, Hwang5}. Although our present aim is to validate the choice of de-Sitter expansion, which relates the potential and the coupling parameter uniquely, however, for the sake of completeness we briefly review the work of Hwang regarding quantum perturbation, and present the expression for the power spectrum in the generalized gravity under consideration. \\

\noindent
As usual, we consider a spatially flat, homogeneous and isotropic Robertson-Walker metric together with the most general scalar-type and tensor-type space-time dependent perturbations as,
\be \label{3.37} ds^2=-a^2\left(1+2\alpha\right)d\eta^2-a^2\left(\beta_{,\alpha}+{B}_{\alpha}\right)d\eta dx^{\alpha}+a^2\left[g_{\alpha\beta}^{(3)}\left(1+2\varphi \right)
+2\gamma_{,\alpha |\beta}+2C_{\alpha |\beta}+2C_{\alpha\beta}\right] dx^{\alpha}dx^{\beta},\ee
where, $dt\equiv ad\eta$, $\eta$ being the conformal time. While, $\alpha(\textbf{x},t)$, $\beta(\textbf{x},t)$, $\varphi(\textbf{x},t)$ and $\gamma(\textbf{x},t)$ characterize the scalar-type perturbations, $B_{\alpha}(\textbf{x},t)$ and $C_{\alpha}(\textbf{x},t)$ are trace-free $(B_{|\alpha}^{\alpha} =0 = C_{|\alpha}^{\alpha})$ and correspond to the vector-type perturbations. Finally, $C_{\alpha\beta}(\textbf{x},t)$ is transverse and trace-free $(C_{\alpha |\beta}^{\beta}=0=C_{\alpha}^{\alpha})$ and corresponds to the tensor-type perturbation. Indices are based on $g_{\alpha\beta}^{(3)}$ as the metric and the vertical bar (`$|$') indicates a covariant derivative. Now, we decompose the energy-momentum tensor along with the scalar field by introducing perturbation as, $T_b^a(\textbf{x},t)=\bar{T}_b^a(t)+\delta T_b^a(\textbf{x},t)$, $\phi(\textbf{x},t)=\bar\phi(t)+\delta\phi(\textbf{x},t)$, and $F =\bar F+\delta F$. In these expressions an over-bar indicates a background ordered quantity and it will be omitted unless necessary. At this stage, we introduce a gauge invariant combination as,

\be\label{3.38}\delta\phi_{\varphi}=\delta\phi-\frac{\dot\phi}{H}\varphi_{\delta\phi},~~~{\delta F\over \dot F} = {\delta \phi\over \dot\phi},\ee
where $H = {\dot a\over a}$ is the Hubble parameter, and since $\delta F$ is related to $\delta\phi$, so one can use either as the representative. Note that $\delta\phi_{\varphi}$ becomes $\delta\phi$ in the uniform-curvature gauge which takes $\phi \equiv 0$ as the gauge condition. The perturbed action in a unified form may be expressed as \cite{Hwang2, Hwang8},

\be\label{3.39} \delta S=\frac{1}{2}\int a^3 \Theta\left\{\dot{\delta\phi_{\varphi}}^2-\frac{1}{a^2}\delta\phi_{\varphi}^
{|\nu}\delta\phi_{\varphi,\nu} +\frac{1}{a^3 \Theta}\frac{H}{\dot\phi}{\left[a^3\Theta {\left(\frac{\dot\phi}{H}\right)}^{.} \right]}^{.}\delta
\phi_{\varphi}^2\right\}dtd^3x,\ee
where, $\Theta=\frac{1+\frac{3\dot F^2}{2\dot\phi^2 F}}{\left(1+\frac{\dot F}{2HF}\right)^2}$, makes all the difference between generalized theory of gravity under consideration and General Theory of Relativity with a minimally coupled scalar field, for which $\Theta = 1$. The equation of motion of $\delta\phi_{\varphi}$ then takes the following form,

\be\label{3.40} \ddot{\delta\phi_{\varphi}}+\frac{(a^3\Theta\dot)}{a^3\Theta}\dot{\delta\phi_{\varphi}}
-\left\{\frac{\nabla^{(3)2}}{a^2} + \frac{1}{a^3\Theta}\frac{H}{\dot\phi} {\left[a^3\Theta {\left(\frac{\dot\phi}{H}\right)}^{.} \right]}^{.}
\right\}{\delta\phi_{\varphi}}=0. \ee
Now, the perturbed form of the  background equation (\ref{3.24}) can be expressed \footnote{Here, $\phi=\bar\phi+\delta\phi_{\varphi}$ and $\dot\phi=\phi_{,i}u^{i}.$ So that $\dot\phi = (\bar\phi+\delta\phi_{\varphi})_{,i}(\bar{u}^i+\delta{u^i})$} as,

\be\label{3.41} \delta\ddot\phi_{\varphi}+3H\delta\dot\phi_{\varphi}-\frac{\nabla^{(3)2}}{a^2} +\frac{1}{2}\left(2V''-\mathrm{f}''\right)
\delta\phi_{\varphi}=0,\ee
where, $\nabla^{(3)2}$ is the Laplacian based on the comoving part of the background three-space metric. Equation (\ref{3.40}) reduces to the perturbed form (\ref{3.41}) of the background metric equation (\ref{3.24}) under the following conditions.
\begin{eqnarray}
                    \label{3.42} \frac{(a^3\Theta\dot)}{a^3\Theta}=3H \\
                   \label{3.43} \frac{1}{a^3\Theta}\frac{H}{\dot\phi} {\left[a^3\Theta {\left(\frac{\dot\phi}{H}\right)}^{.}
                   \right]}^{.}&=& \frac{1}{2}\left(2V'' -\mathrm{f}''\right)=\theta_0,
                  \end{eqnarray}
where, $\theta_0$ is a constant. Now, the above pair of conditions \eqref{3.42} and \eqref{3.43} lead to

\begin{eqnarray}\label{3.44a}\Theta=\text{constant},\\
\label{3.44b} 2V' - \mathrm{f}' = 2\theta_0\phi,\end{eqnarray}
where we have set the constant of integration in \eqref{3.44b} to zero. In the process we find exactly the same relationship \eqref{Rel1} between the generalized parameter $\mathrm{f}(\phi, R)$ and the potential $V(\phi)$, under the choice $\theta_0 = 2H^2$, as was found while seeking inflationary solution of the classical field equations in the de-Sitter exponential form. This fact proves without ambiguity that the model under consideration \eqref{A2}, allows de-Sitter expansion in the inflationary regime rather than the power law expansion. Further, it also proves that the technique followed to find gravitational perturbation for non-standard models is legitimate. This is another important finding of the present work.\\

Now, introducing new variables, $z\equiv a\sqrt{\Theta}\frac{\dot\phi}{H}$ and ${v}\equiv \sqrt{\Theta}a\delta\phi_{\varphi}$, one can express the above equation (\ref{3.40}) in terms of $\varphi_{\delta\phi}$ as

\be\label{3.45}v_{,\eta\eta}+\left(k^2-\frac{z_{,\eta\eta}}{z}\right)v=0.\ee
In the above, comma denotes ordinary derivative. The asymptotic solutions of equation (\ref{3.45}) in the large-scale $(k^2\ll \frac{z_{,\eta\eta}}{z})$ and small-scale $(k^2\gg \frac{z_{,\eta\eta}}{z})$ limits are,
\be\label{3.46}\delta \phi_{\varphi}(\textbf{x},t)=-\frac{\dot\phi}{H}\left[{C(\textbf{x})}-{D(\textbf{x})}
\int_0^t{\frac{1}{a^3\Theta}\frac{H^2}{\dot\phi^2}dt}\right],\ee
\be\label{3.47}\delta \phi_{\varphi}(\textbf{k},\eta)=\frac{1}{a\sqrt{2{k}}}\left[c_1(\textbf{k})
e^{i{k}\eta} +c_1(\textbf{k})e^{-i{k}\eta}\right]\frac{1}{\sqrt{\Theta}},  \ee
where, $C(\textbf{x})$, $D(\textbf{x})$ are integration constants of the relatively growing and decaying modes and $c_1(\textbf{k})$, $c_2(\textbf{k})$ are arbitrary integration constants, respectively. Further, as $\frac{z_{,\eta\eta}}{z}=\frac{n}{\eta^2}$,  where $n=$ constant, the solution of equation (\ref{3.40}) can be expressed in the following form \cite{Martin, Lyth, TTN},
\be\label{3.48} \delta \phi_{\varphi \textbf{k}}(\eta)=\frac{\sqrt{\pi |\eta|}}{2a}\left[C_1(\textbf{k}) \mathbf{H}_{\nu}^{(1)}
(\textbf{k}|\eta|)+D_1(\textbf{k})\mathbf{H}_{\nu}^{(2)}(\textbf{k}|\eta|) \right]\frac{1}{\sqrt{\Theta}}, \ee
where $C_1$ and $D_1$ are integration constants, $\nu=\sqrt{n+\frac{1}{4}}$ and the Hankel functions $\mathbf{H}_{\nu}^{(1,2)}(x)\rightarrow \pm \frac{\Gamma(\nu)(\frac{2}{x})^{\nu}}{i\pi}.$

\subsection{Perturbative semiclassical approach}

In the post Planck era gravity should be treated classically, while the matter fields still behave quantum mechanically. 'Quantum Field Theory in Curved Space-Time' (QFT in CST), treats gravity classically while the energy-momentum tensor is quantized. For General Theory of Relativity, it is expressed as $G_{\mu\nu} = <\widehat{T}_{\mu\nu}>$, where the quantum operator $<\widehat{T}_{\mu\nu}>$ is a suitably renormalized expectation value. The perturbative semiclassical approximation under consideration in this subsection, is quite different from QFT in CST. In this technique, the perturbed parts of the metric and matter fields are treated as quantum mechanical operators, keeping the background parts classical \cite{Hwang5, Hwang6, Hwang7}. Since, we derive the perturbation spectra generated from the quantum fluctuations in an early scalar field dominated accelerating (inflation) phase, therefore this approach appears to be more legitimate than QFT in CST. However, we shall use the result of QFT in CST to choose appropriate vacuum.\\

Instead of the classical decomposition, we are therefore required to replace the perturbed order variables with the quantum operator in Heisenberg representation of $\delta \hat\phi(\mathbf{x},t)$ as,

\be\phi(\mathbf{x},t)=\phi(t)+ {\delta} \hat\phi(\mathbf{x},t);~~~\delta\hat{\phi}_{\varphi} = \delta\hat\phi - {\dot \phi\over H}{\hat\varphi}.\ee
Note that the background order quantities are considered as classical variables. Now, in the flat space under consideration it is possible to expand $\delta\hat{\phi}(\mathbf{x},t)$ in the mode expression as

\be\delta\hat{\phi}(\mathbf{x},t)=\int{\frac{d^3k}{(2\pi)^{\frac{3}{2}}}\left[\hat a_{\textbf{k}}\delta\phi_{\textbf{k}}(t)e^{i\textbf{k}.\textbf{x}}
+\hat a_{\textbf{k}}^{\dag}\delta\phi_{\textbf{k}}^*(t)e^{-i\textbf{k}.\textbf{x}}\right]},\ee
where, the annihilation and creation operators, $\hat a_{\textbf{k}}$ and $\hat a_{\textbf{k}}^\dag$ satisfy standard commutation relation:

\be\label{cr}[\hat a_{\textbf{k}},\hat a_{\textbf{k}'}]=0=[\hat a_{\textbf{k}}^\dag,\hat a_{\textbf{k}'}^\dag],~~ \mathrm{and},~~ [\hat a_{\textbf{k}},\hat a_{\textbf{k}'}^\dag] =\delta^3(\textbf{k}-\textbf{k}'),\ee
while $\delta\phi_{\textbf{k}}(t)$ is the mode function, a complex solution of the classical mode evolution equation \eqref{3.40}. Equal time commutation relation must also hold for $\delta\hat{\phi}(\mathbf{x},t)$ and its canonical conjugate momentum, which reads

\be\label{etcr} \left[\delta\hat{\phi}_{\varphi}(\mathbf{x}, t), \delta\dot{\hat{\phi}}_{\varphi}(\mathbf{x'}, t)\right] = {i\over a^3 \Theta} \delta^3(\mathbf{x}-\mathbf{x'}).\ee
In order that the commutation relations \eqref{cr} and \eqref{etcr} hold simultaneously, the mode function should satisfy the Wronskian condition,

\be\label{wrons} \delta\phi_{\varphi\mathbf{k}}\delta\dot{\phi}^*_{\varphi\mathbf{k}}-\delta\phi^*_{\varphi\mathbf{k}}\delta\dot{\phi}_{\varphi\mathbf{k}}={i\over a^3 \Theta}.\ee
At this stage assuming ${z_{,\eta\eta}\over z} = {n\over \eta}$, the implication of which has been discussed in details by Hwang in \cite{Hwang7}, following solution is found

\be \label{solution}\delta\phi_{\varphi\mathbf{k}}(\eta) = {\sqrt \pi|\eta|\over 2 a}\left[c_1(\mathbf{k})\mathrm{H}^{(1)}_\nu(k\eta) + c_2(\mathbf{k})\mathrm{H}^{(2)}_\nu(k\eta)\right]{1\over\Theta},\ee
where, $\nu = \sqrt{n + {1\over 4}}$. Coefficients $c_1(\mathbf{k})$ and $c_2(\mathbf{k})$ are arbitrary functions of $\mathbf{k}$, which are normalized in accordance with \eqref{wrons} as

\be  |c_1(\mathbf{k})|^2 - |c_2(\mathbf{k})|^2 = 1.\ee
Clearly imposing quantum condition in \eqref{wrons} does not completely fix the coefficients. The remaining freedom depends on the choice of vacuum state. Here, we insert the result of QFT in CST to choose adiabatic vacuum, also known as the Bunch-Davies vacuum in de-Sitter space, which fixes $c_1(\mathbf{k}) = 0$ and $c_2(\mathbf{k}) = 1$, corresponding to positive frequency solution in Minkowski space limit. The power spectrum therefore takes the form,

\be\mathcal{P}_{\delta\hat{\phi}_\varphi}(k,t)=\frac{k^3}{2\pi^2}|\delta\phi_{\varphi\textbf k}(t)|^2 =\frac{k^3}{2\pi^2}\int<\delta\hat{\phi}_{\varphi}(\textbf{x}+\textbf{r},t)
\delta\hat{\phi}_{\varphi}(\textbf x, t)>_{vac}e^{-i\textbf{k}.\textbf r} d^3\textbf r,\ee
where, $<  >_{\text{vac}}\equiv <\text{vac}| |\text{vac}>$ is a vacuum expectation value and $a_{\textbf k}|\text{vac}>\equiv 0$ for every $\textbf k$. Now, assuming adiabatic vacuum, two point function $\textbf{G}(x',x'')$ which is defined as,

\be\begin{split}&\textbf{G}(x',x'')\equiv <\delta\hat{\phi}_\varphi(x')\delta\hat{\phi}_\varphi(x'')>_{vac}=
\int{\frac{d^3k}{(2\pi)^3}}e^{i\textbf{k}.(\textbf{x}'-\textbf{x}'')}\delta\phi_{\textbf k}(t')\delta\phi_{\textbf k}^*(t'')\\&
=\int_0^\infty{\frac{k^2dk}{2\pi^2}}J_0({{k}|\textbf{x}'-\textbf{x}''|})
\delta\phi_{\textbf k}(t')\delta\phi_{\textbf k}^*(t'')\end{split},\ee
where, $x\equiv(\textbf{x}, t)$, takes the form in the exponential expansion case ($\eta = -{1\over aH}$),

\be \textbf{G}(x',x'') = \frac{({1\over 4}-\nu^2)\sec(\pi\nu)}{16\pi a' a'' \eta'\eta''}\times \Gamma\left({3\over 2}+\nu,~{3\over 2}-\nu;~2;~1;~1+\frac{(\eta'-\eta'')^2-(\mathbf{x'} - \mathbf{x''})^2}{4\eta'\eta''}\right){1\over \sqrt{\Theta'\Theta''}},\ee
which is valid for $\nu < {3\over 2}$. At the equal time, the vacuum expectation value reads

\be <\hat{\delta\phi}(\textbf{x}+\textbf{r},t)\hat{\delta\phi}(\textbf x, t)>_{vac}=\int{\frac{d^3k}{(2\pi)^3}}e^{i\textbf{k}.\textbf r} |\delta\phi_{\textbf k}(t)|^2 =\int_0^\infty{\mathcal{P}_{\hat{\delta\phi}}(k,t)J_0(kr) d\ln k}.\ee
In the small scale limit the solution \eqref{solution} takes the form

\be \delta\phi_{\varphi\mathbf{k}}(\eta) = {1\over a\sqrt{2k\Theta}}e^{-ik\eta +i(\nu+{1\over 2}){\pi\over 2}}.\ee
On the contrary, in the large scale limit we can write from (\ref{solution})

\be \delta\phi_{\varphi\mathbf{k}}(\eta) = i{\sqrt|\eta| \Gamma(\nu) \over 2 a \sqrt{\pi\Theta}}\left({k|\eta|\over 2}\right)^{-\nu},\ee
and the power spectrum reads

\be\label{Pp}\mathcal{P}_{\delta\hat{\phi}_\varphi}(k,\eta)=\frac{\Gamma(\nu)}{\pi^{3\over 2}a|\eta|}\left({k|\eta|\over 2}\right)^{{3\over 2}-\nu}
\frac{1}{\sqrt\Theta}.\ee
Since $\Theta$ is a constant in view of \eqref{3.44a}, so the power spectrum may deviate only by a constant factor from the minimally coupled case for which $\Theta = 1$.

\section{Concluding remarks}

In the absence of a complete theory of gravity, quantum cosmology is studied to perceive the evolution of the universe in the Planck regime, when gravity is quantized. Shortly after the Plank's era, gravity becomes classical, while the matter fields still remain quantized. In this era, the universe went through an inflationary phase. The seed of perturbation grows, which finally leads to the structures we observe. In this manuscript, we have found the wave-function of a model universe which contains scalar curvature squared term. The effective Hamiltonian operator is hermitian and standard quantum mechanical probability interpretation holds, which prove that the wave-function is stable and well behaved. The model universe transits smoothly to a de-Sitter type inflationary phase, and inflationary parameters have been found to be very much within the observational limit. The perturbation analysis shows that only de-Sitter solution is allowed, since it relates the coupling parameter and the potential in the same manner as de-Sitter solution to the classical field equations does. Perturbative semiclassical approximation method has been pursued in which the perturbed parts of the metric and matter fields are treated as quantum mechanical operators, keeping the background parts classical. In the process, we derive the perturbation spectra generated from the quantum fluctuations in an early scalar field dominated de-Sitter phase. The power-spectrum deviates only by a constant term from the minimally coupled case.\\

Regarding the canonical quantization scheme followed here and the results obtained in the present manuscript, there are a few important issues to discuss. We have at least in one case (G-B-dilatonic coupled gravity in the presence of scalar curvature squared term) proved that the known standard techniques (\cite{ostro, dirac, horo}) regarding canonical formulation of higher order theory of gravity do not produce a viable quantum theory, since these techniques don't reveal correct classical analogue under semi-classical approximation \cite{mod6}. It is therefore legitimate to follow the technique of canonical formulation, which is well behaved in all the cases studied so far. Here, we have considered a modified (with higher order term) non-minimally coupled scalar-tensor gravitational action, and aimed at studying the very early universe. Canonical formulation in isotropic and homogeneous background has been performed following modified Horowitz' formalism, which has been extensively applied earlier in different situations. The most important result as already mentioned may be summarized in the following manner. The forms of the coupling parameter $f(\phi)$ and the potential $V(\phi)$ obtained following perturbative analysis, dictates that the classical field equations must admit de-Sitter type exponential expansion in the inflationary regime. The wave function of the universe also admits the same de-Sitter type expansion under semiclassical approximation. This means that the very early universe smoothly transits to exponential expansion (inflationary) phase. This result proves overall consistency of the present study. Finally, although we are neither the proponents of multiverse theory nor do we oppose it as a topic of theological discourse, it should be mentioned that a stable quantum dynamics in the Trans-Planckian era doesn't in any way rule out the possibility of multiverse, which may have been created during the next inflationary phase.. The potential we found is sufficiently flat when the scalar field is large enough, and therefore the possibility of multiverse is nascent in the quantum fluctuation of such a field.

\appendix

\section{Hermiticity of the effective Hamiltonian $\hat{H_e}$}
To prove that the effective Hamiltonian is hermitian, let us split it as
\be \hat{H_e} = \hat{H_1} + \hat{H_2} + \hat{H_3} + \hat{V_e},\ee
where,
\be\begin{split} &\hat{H_1}= -\frac{\hbar^2}{54B}\left(\frac{1}{x}\frac{\partial^2}{\partial x^2}+\frac{n}{x^2}\frac{\partial}{\partial x}\right),\\&
\hat{H_2} = - \frac{\hbar^2}{3x \sigma^{\frac{4}{3}}}\frac{\partial^2}{\partial\phi^2},\\&
\hat{H_3}=\frac{i\hbar}{3\sigma}\left(\frac{\phi^3+6f_1}{\phi^2}\right)\frac{\partial}{\partial\phi}+\frac{i\hbar}{6\sigma}\left(\frac{\phi^3-12f_1}{\phi^3} \right).\end{split}\ee
Let us first take the very first term.
\be \int(\hat{H_1}\psi^*)\psi dx = -\frac{\hbar^2}{54B}\int\left(\frac{1}{x}\frac{\partial^2 \psi^*}{\partial x^2}+\frac{n}{x^2}\frac{\partial\psi^*}{\partial x}\right)\psi dx = -\frac{\hbar^2}{54B}\int\left(\frac{\psi}{x}\frac{\partial^2 \psi^*}{\partial x^2}+\frac{n\psi}{x^2}\frac{\partial\psi^*}{\partial x}\right) dx\ee
Under integration by parts twice and dropping the first (integrated out) term due to fall-of condition, we obtain,
\be \begin{split}\int(\hat{H_1}\psi^*)\psi dx &=-\frac{\hbar^2}{54B}\int \psi^*\left[{1\over x}{\partial^2\psi\over\partial x^2} - {\left(2+n\over x^2\right)}{\partial\psi\over\partial x} + {2(n+1)\over x^3}\psi\right]dx \\&
= -\frac{\hbar^2}{54B}\int\psi^*\left[{1\over x}{\partial^2\psi\over \partial x^2} -{1\over x^2}{\partial\psi\over\partial x}\right]dx = \int \psi^* \hat H_1 \psi dx, \end{split}\ee
under the choice $n = -1$. Thus, $\hat H_1$ is hermitian, under a particular choice of operator ordering parameter $n = -1$. It is trivial to prove that $\hat H_2$ is hermitian. We therefore turn our attention to $\hat H_3$.

\be \begin{split}\int(\hat{H_3}\psi^*)\psi d\phi &=-{i\hbar\over 3\sigma}\int\left(\phi+{6f_1\over\phi^2}\right){\partial\psi^*\over \partial\phi}\psi d\phi -{i\hbar\over 6\sigma}\int\left(1-{12f_1\over \phi^3}\right)\psi^*\psi d\phi.\end{split}\ee
Again under integration by parts and dropping the integrated out terms due to fall-of condition, it is possible to arrive at

\be \begin{split}\int(\hat{H_3}\psi^*)\psi d\phi &= {i\hbar\over 3\sigma}\int\psi^*\left(\phi+{6f_1\over\phi^2}\right){\partial\psi\over \partial\phi} d\phi +{i\hbar\over 6\sigma}\int\left(1-{12f_1\over \phi^3}\right)\psi^*\psi d\phi = \int \psi^*\hat{H_3}\psi d\phi,\end{split}\ee
and the Weyl symmetric ordering performed between $f'(\phi)$ and $p_{\phi}$ turns out $\hat{H_3}$ to be hermitian operator, and thus the effective Hamiltonian ($\hat{H_e}$) as such is hermitian operator.

\section{Canonical formulation following standard techniques}

While in the methodology of canonical formulation of higher order theory of gravity adopted here, we insist on fixing the three-space metric $h_{ij}$ and the Ricci scalar $R$ at the boundary, the standard methods keep $h_{ij}$ and the extrinsic curvature tensor $K_{ij}$ fixed at the boundary, instead. In the process, all the boundary terms, obtained under integration by parts vanish and there is no need to supplement the action by boundary terms as presented in action \eqref{2.3}. As a result it is legitimate to start with action \eqref{A2}. In the appendix, we show that canonical formulation following some standard methods lead to a unique Hamiltonian. However it is different from and is not canonically related to the one \eqref{2.15} obtained in the manuscript.

\subsection{Ostrogradski's formalism}

The Hamiltonian formulation of theories with higher derivatives was first developed by Ostrogradski more than one and half a century ago \cite{ostro}, which gives special treatment to the highest derivatives of the original Lagrangian, so that the initial higher-order regular system be reduced to a first-order system. Ostrogradski's theorem can be stated as follows: If the higher order time derivative Lagrangian is non-degenerate, there is at least one linear instability in the Hamiltonian of this system. Basically, the method \cite{ostro} is adopted only to systems described by regular Lagrangian, that is, a Lagrange function for which the associated Hessian matrix formed with respect to the highest-order time derivatives has a nonzero determinant. The underlying idea of this method and of its subsequent generalizations consist in introducing, besides the original configuration variables, a new set of coordinates that encompasses each of the successive time derivatives of the original Lagrangian coordinates so that the initial higher-order regular system be reduced to a first-order system. In Ostrogradski's formalism \cite{ostroappl1, ostroappl2}, if a Lagrangian contains maximum order of $m$-th time derivatives of the generalized coordinate $q_i$,
in the form $L = L(q_i, \dot q_i, \ddot q_i, \dddot q_i...., ^m{q_i})$, where, $^m{q_i} = ({d\over dt})^m$, then one should choose $m$ independent variables
$(q_i, \dot q_i, \ddot q_i, \dddot q_i...., ^{m-1}{q_i})$ and corresponding m generalized momenta $p_{i,0}, p_{i,1}, p_{i,2}...p_{i,m-1}$ according to the
recurrence relations, $p_{i, m-1} = {\partial L\over \partial q_i^m}$, and $p_{i, n-1} = {\partial L\over \partial q_i^n} - \dot p_{i,n}$, for $n = 1, 2, ...m-1$.
With these new independent coordinates and their corresponding momenta, the Legendre transformation to express the phase-space Hamiltonian reads
\be H_{\mathcal O} = \sum_{i=1}^N\sum_{\alpha = }^{m-1}\dot q_i^{\alpha}p_{i,\alpha} - L.\ee
Starting with action \eqref{2.6}, apart from the supplementary boundary terms, we therefore, for the present purpose require two variable $z$ and $x$, where $x = {\dot z\over N} = -2 K_{ij}$. The action (\ref{2.6}) may therefore be expressed as,

\be\label{ao}  A=\int\Big[3f\sqrt z\left(\dot x+2kN\right)+\frac{9B}{N\sqrt z}\left(\dot x+2kN\right)^2+ z^{\frac{3}{2}}
\Big(\frac{\dot\phi^2}{2N}-VN\Big)\Big] dt. \ee
Since, the Hessian determinant vanishes, so the corresponding Lagrangian is singular, and as a result, canonical formulation following Ostrogradski's formalism is not possible. However, if we make a gauge fixing $N=1$ a-priori, the Hessian is non-vanishing and Ostrogradski's technique may be pursued. The canonical
momenta in that case are,
\be p_x=3f\sqrt{z}+\frac{18B}{\sqrt{z}}\left({\dot x} + 2k  \right);~ p_z=\frac{\partial L}{\partial\dot z}-\dot {p_x}
=-\frac{18B\ddot x}{\sqrt{z}}+\frac{9Bx}{z^{\frac{3}{2}}}\left({\dot x} + 2k  \right)-3f'\dot\phi\sqrt{z}-\frac{3fx}{2\sqrt{z}};~ p_{\phi}=
{\dot\phi z^{\frac{3}{2}}\over N}.\ee
In terms of phase space variables the Hamiltonian reads,
\be \label{ho} \mathcal{H_{O}} = xp_z+\frac{\sqrt z}{36B}p_x^2-2kp_x-\frac{fzp_x}{6B}+\frac{f^2z^{\frac{3}{2}}}{4B}+
\frac{p_{\phi}^2}{2z^{\frac{3}{2}}}+Vz^{\frac{3}{2}}.\ee

\subsection{Dirac's constraint analysis}

As already mentioned, for arbitrary $N$, the Hessian vanishes, and so canonical formulation may be pursued following Dirac's constraint analysis \cite{dirac}.
We introduce $(\frac{{\dot z}}{N}-  x)=0$ through Lagrange multiplier $\lambda$, and the point Lagrangian corresponding to action \eqref{2.6} therefore reads,
\be\begin{split}\label{dl} L&=3f\sqrt z\left(\dot x+2kN\right)+\frac{9B}{N\sqrt z}\left(\dot x+2kN\right)^2+z^{\frac{3}{2}}
\left(\frac{1}{2N}\dot\phi^2-VN\right)+\lambda\left(\frac{{\dot z}}{N}- x\right).\end{split}\ee
Canonical momenta are
\be\label{dm} p_x=3f\sqrt{z}+\frac{18B}{N\sqrt z}(\dot x+2kN),\;\ p_z=\frac{\lambda}{N},\;\ p_{\phi}={{\dot\phi}z^{\frac{3}{2}}\over N},
 \;\ p_N=0,\;\ p_{\lambda}=0\ee
The constraint Hamiltonian therefore is
\be\label{dch} H_{c}=\dot xp_{ x}+\dot zp_z+\dot \phi p_\phi+\dot Np_N+\dot\lambda p_{\lambda}-L\ee
Clearly we require three primary constraints involving Lagrange multiplier or its conjugate viz,
$\phi_1=p_z-{\lambda},~~ \phi_2=p_{\lambda},~~ \phi_3=p_N$.
Since, the lapse function $N$ is non-dynamical, so we have safely considered the associated constraint to vanish strongly. The first two constraints can now
be harmlessly substituted into the modified primary Hamiltonian, which takes the form,
\be\begin{split}\label{dmph}~~~~~ H_{p1}&= \frac{N\sqrt z}{36B}p_x^2-2kNp_x-\frac{fzNp_x}{6B}+\frac{Nf^2z^{\frac{3}{2}}}{4B}+\frac{Np_{\phi}^2}{2z^{\frac{3}{2}}}
 +VNz^{\frac{3}{2}}+\lambda x+u_1\big(Np_z-{\lambda}\big)+u_2p_{\lambda}.\end{split}\ee
Here $u_1, u_2$  are the Lagrange multipliers and the Poisson bracket $\{x,p_x\}=\{z,p_z\}=\{\lambda,p_{\lambda}\}=1$, hold. Now constraint should remain preserved in time, which are exhibited in the Poisson brackets $\{\phi_i,H_{p1}\}$ viz,
\be\begin{split}\dot\phi_1 &=\{\phi_1,H_{p1}\}=-N\frac{\partial H_{p1}}{\partial z}-u_2+
\Sigma_{i=1}^2\phi_i\{\phi_1,u_i\}\end{split}.\ee
\be\dot\phi_2=\{\phi_2,H_{p1}\}=-x+u_1+\Sigma_{i=1}^2\phi_i\{\phi_2,u_i\} \ee
Consequently, all the Poisson bracket relations vanish weakly if we set,
\be\begin{split}u_2&= -N\frac{\partial H_{p1}}{\partial z}\end{split}\;\;\; \mathrm{ and}\;\;\;u_1=x.\ee
Since, under the condition $\dot\phi_1$ and $\dot\phi_2$ should vanish weakly, explicit forms of $u_2$ and $u_1$ are found, so the modified primary Hamiltonian is,
\be\begin{split}\label{dmph2}H_{p2}&= \frac{N\sqrt z}{36B}p_x^2-2kNp_x-\frac{fzNp_x}{6B}+\frac{Nf^2z^{\frac{3}{2}}}{4B}+ \frac{Np_{\phi}^2}
{2z^{\frac{3}{2}}}+VNz^{\frac{3}{2}} +Nxp_z -N\frac{\partial H_{p1}}{\partial z}p_{\lambda}.\end{split}\ee
Now under the same condition that constraint should remain preserved in time, $p_{\lambda}$ vanishes trivially. Therefore, the Hamiltonian finally takes the form,
\be\label{hd} \begin{split} H_D&=N\left[xp_z + \frac{\sqrt z}{36B}p_x^2-2kp_x-\frac{fzp_x}{6B}+\frac{f^2z^{\frac{3}{2}}}{4B}+
\frac{p_{\phi}^2}{2z^{\frac{3}{2}}}+Vz^{\frac{3}{2}}\right]=N\mathcal{H_D}.\end{split}\ee
Clearly, $\mathcal{H_D}$ is no different from $\mathcal{H_O}$.

\subsection{Horowitz' formalism}

Instead of choosing the basic variables as $\{h_{ij}, Q_{ij}\}$ together with their conjugate momenta, $\{p^{ij}, \Pi^{ij}\}$, (where, $Q^{ij}
= -{\partial{\mathcal{L}}\over\partial({\mathcal{L}}_n K_{ij})}$), as considered by Boulware \cite{Boul}, Horowitz \cite{horo} took another set of variables viz.
$\{h_{ij}, K_{ij}\}$ together with their conjugate momenta, $\{p^{ij}, P^{ij}\}$. Although, the resulting formalism is completely equivalent to \cite{Boul},
the latter appears to be much convenient to handle. The quantized version (the modified Wheeler-deWitt equation) in this formalism, corresponding to the positive definite action, resembles with Schr\"odinger equation, where the internal parameter, viz. the three metric $h_{ij}$ plays the role of time. Horowitz \cite{horo} argued against supplementary boundary terms and insisted on keeping $h_{ij}~\mathrm{and}~K_{ij}$ fixed at the boundary. So that higher order theory is devoid of supplementary boundary terms. However, treating $K_{ij}$ as a variable from the beginning, requires to vary the action with respect to $K_{ij}$ as well, together with $h_{ij}$, since both are treated on the same footing. This restricts classical solutions by and large. Therefore, in order to obtain the canonical structure, Horowitz, started with an auxiliary variable $Q_{ij}$, which is found by varying the action with respect to the highest derivative of the field variables present in the action. The Hamiltonian so obtained, was finally expressed in terms of the basic variables $\{h_{ij}, K_{ij}; p^{ij}, \Pi^{ij}\}$, following canonical transformation. Let us start with the action (\ref{2.6}) apart from the supplementary boundary terms to explore the situation. Introducing the auxiliary variable

\be\label{Q} Q={\partial L\over \partial\ddot z} =\frac{3f\sqrt{z}}{N}+ \frac{18B}{N\sqrt{z}}\Big(\frac{\ddot z}{N^2}-
\frac{\dot N\dot z}{N^3}+2k\Big)\ee
straight into the action (\ref{2.6}) as
\be\begin{split}\label{7}
A &= \int\Big[Q\ddot z-\frac{N^3\sqrt{z}Q^2}{36B}-\frac{Nf^2 z^{\frac{3}{2}}}{4B}+\frac{fN^2 Qz}{6B}+2kQN^2 -\frac{\dot N\dot z Q}{N} +
\frac{z^{\frac{3}{2}}\dot\phi^2}{2N}-VNz^{\frac{3}{2}}\Big]dt,\end{split} \ee
and after integration by parts, resulting canonical action reads
\be\begin{split}\label{71}
A &= \int\Big[-\dot Q\dot z-\frac{N^3\sqrt{z}Q^2}{36B}-\frac{Nf^2 z^{\frac{3}{2}}}{4B}+\frac{fN^2 Qz}{6B}+2kQN^2 -\frac{\dot N\dot z Q}{N} +
\frac{z^{\frac{3}{2}}\dot\phi^2}{2N}-VNz^{\frac{3}{2}} \Big]dt.\end{split} \ee
Now in view of the canonical momenta
 \be\label{ph} p_z =-\dot Q-{\dot N Q\over N};\;\;\; p_\phi = {{z^{\frac{3}{2}}\dot\phi}\over N};\;\;\; p_Q =
  -\dot z;\;\;\;p_N=-\frac{\dot z Q}{N}
\ee
the Hamiltonian in phase space variables is readily obtained as,
\be\begin{split}\label{hh1}
 H_H &=-p_Q p_z+\frac{N^3\sqrt{z}Q^2}{36B}-2kQN^2-\frac{fN^2 Qz}{6B}+\frac{Nf^2 z^{\frac{3}{2}}}{4B} +\frac{Np^2_{\phi}}{2z^{\frac{3}{2}}}
+VNz^{\frac{3}{2}}.\end{split}\ee
It is now required to express the Hamiltonian in terms of basic variables ($z$ and $x =\frac{\dot z}{N}$), instead of auxiliary variable ($z$ and $Q$).
Since, $p_Q =-\dot z=-Nx$ and $Q =\frac{p_x}{N}$, one therefore is required to make the following canonical transformation with the replacements of, $p_Q$
by $-Nx$ and $Q$ by $\frac{p_x}{N}$ in the Hamiltonian (\ref{hh1}), which therefore finally results in,
\be\begin{split}\label{hh}
 H_H &=N\left[x p_z+\frac{\sqrt z}{36B}p_x^2-2kp_x-\frac{f z p_x}{6B}+\frac{f^2z^{\frac{3}{2}}}{4B}+\frac{p_{\phi}^2}{2z^{\frac{3}{2}}}+
 Vz^{\frac{3}{2}}\right]
 =N\mathcal{H_H}.\end{split}\ee
One can readily observe that the Hamiltonian obtained following the above three formalism, viz.,  Ostrogradski \eqref{ho}, Dirac \eqref{hd} and Horowitz \eqref{hh})
are the same $\mathcal{H_O = H_D = H_H}$. However, the Hamiltonian
$\mathcal H$ \eqref{2.15} is different from $\mathcal H_O$ \eqref{ho}, $\mathcal H_D$ \eqref{hd} or $\mathcal H_H$ \eqref{hh}. Although the set of
transformations
\be\begin{split}
&z=Z,~~p_z= P_Z-18B\frac{kX}{Z^{\frac{3}{2}}}+\frac{3f X}{2\sqrt{Z}}; ~~~~~x = X,~~p_x=P_X+36B \frac{k}{\sqrt{Z}}+3f\sqrt{Z};\\ &
\phi=\Phi,~~p_{\phi}={P_{\Phi}+{3f' X\sqrt Z}},
\end{split}\ee
relates $\mathcal H$ with the others, nevertheless, such transformations are not canonical. This establishes the fact that different boundary conditions
lead to different Hamiltonian in general, and they are not related under canonical transformations. The reason for following modified Horowitz' formalism has been discussed in the introduction and in the concluding remarks.

\end{document}